\documentclass[aps,twocolumn,showpacs,prd,nofootinbib,preprintnumbers]{revtex4}
\usepackage{graphicx}
\usepackage{dcolumn}
\usepackage{bm}
\usepackage{amsmath, amssymb}
\usepackage{color}

\begin{document}
\preprint{EPHOU-12-002}

\title
{\Large\bf Grand Unification of Flavor Mixings}

\author{Naoyuki Haba$^1$}

\author{Ryo Takahashi$^{1,2}$}

\affiliation{
$^{1}$Department of Physics, Faculty of Science, Hokkaido University, Sapporo 
060-0810, Japan}
\affiliation{$^{2}$Department of Physics, Graduate School of Science, Osaka 
University, Toyonaka, Osaka 560-0043, Japan}

\begin{abstract}
An origin of flavor mixings in quark and lepton sectors is still a mystery, and
 a structure of the flavor mixings in lepton sector seems completely different 
from that of quark sector. In this letter, we point out that the flavor mixing 
angles in quark and lepton sectors could be unified at a high energy scale, 
when neutrinos are degenerate. It means that a minimal flavor violation at a 
high energy scale can induce a rich variety of flavor mixings in quark and 
lepton sectors at a low energy scale through quantum corrections. 

\noindent
\end{abstract}
\maketitle

An origin of flavor mixings is one of the most important mystery in the 
elementary particle physics. A structure of flavor mixings in the quark sector 
has been investigated as so-called Cabibbo-Kobayashi-Maskawa (CKM) 
matrix~\cite{Cabibbo:1963yz}. On the other hand, neutrino oscillation 
experiments have revealed that the lepton sector has completely different flavor
 mixings, represented by Pontecorvo-Maki-Nakagawa-Sakata (PMNS) 
matrix~\cite{Maki:1962mu}, in which one large mixing angle $\theta_{12}$, one 
nearly maximal mixing angle $\theta_{23}$~\cite{Nakamura:2010zzi}, and 
non-vanishing $\theta_{13}$ that is pointed out by recent long baseline and 
reactor experiments~\cite{An:2012eh}. Anyhow, both flavor structures seem 
completely different from each other, and this situation motivates us to pursue
 an origin of flavor violation. 

In this letter, we will investigate a possibility that CKM and PMNS flavor 
mixing angles are unified at a high energy scale. This is a kind of 
``grand unification of flavor mixings (GUFM)'', where a minimal flavor violation
 at a high energy scale can induce a rich variety of flavor mixing structures in
 both quark and lepton sectors at a low energy scale. Similar possibility has 
been studied in~\cite{Mohapatra:2003tw,Mohapatra:2005gs,Agarwalla:2006dj}. Such 
possibility in~\cite{Mohapatra:2003tw} has been realized by a radiative 
magnification~\cite{Balaji:2000au}. The idea of radiative magnification has 
originally proposed for the neutrino mixing angles but not for an unification of
 CKM and PMNS mixing angles (see~\cite{8} for radiative magnification models). 
The refs.~\cite{Mohapatra:2003tw,Mohapatra:2005gs,Agarwalla:2006dj} could give 
some typical examples with the radiative magnification which can cause to the 
GUFM. The ref.~\cite{Samanta:2011pd} has applied the GUFM to phenomenological 
discussions, i.e. it has been shown that there is a correlation between lower 
bounds on masses of super-particles and an upper bound on sum of neutrino 
masses. Our purposes of this work are to clarify parameter space and give some 
bounds on physical parameters at a low energy for a realization of the GUFM 
rather than a construction of high energy model, which realizes the GUFM, and 
phenomenological applications. Therefore, we will take a bottom-up approach with
 renormalization group equations (RGEs) and experimentally observed values at a 
low energy as input, which includes the recent update of value of $\theta_{13}$.
 Then, we will show that quantum corrections and degeneracy of neutrino masses 
play crucial roles for the realization of GUFM. As for the mass degeneracy, we 
should remind that only neutrinos can be degenerate among matter fermions.

There is also a similar work to the GUFM, which is a quark-lepton 
similarity~\cite{Hwang:2010wm}. The ref.~\cite{Hwang:2010wm} has pointed out 
that the PMNS matrix at high energy scale can be connected to the CKM matrix by 
a transformation. We will consider a possibility of GUFM without introducing 
such special transformation, i.e. the GUFM will be discussed under the RGEs 
with standard PDG parameterization for both CKM and PMNS matrices.

We take a setup of minimal supersymmetric standard model (MSSM) with Weinberg 
operator~\cite{Weinberg:1979sa}, where Yukawa interactions are given by 
 \begin{eqnarray}
  \mathcal{L}_Y &=& -y_d\overline{Q_L}H_dd_R-y_u\overline{Q_L}H_uu_R
                    -y_e\bar{L}H_de_R \nonumber \\ 
                & & +\kappa(H_uL)(H_uL)+h.c..
 \end{eqnarray}
Here $Q_L$ are the left-handed quarks, $u_R(d_R)$ are the right-handed 
up(down)-type quarks, $e_R$ are the right-handed charged leptons, $H_u(H_d)$ is
 the up(down)-type Higgs, and $y_\ast$ ($\ast=u,d,e$) are the corresponding 
 Yukawa couplings, respectively. The Weinberg operator can be obtained by 
integrating out a heavy particle(s), for example, right-handed neutrinos with 
masses of order $10^{14-16}$ GeV in type I seesaw mechanism~\cite{seesaw}. 
Thus, an effective coupling $\kappa$ is carrying mass dimension $-1$ as (${\cal 
O} (10^{14-16})$ GeV)$^{-1}$. We also utilize PDG 
parameterization~\cite{Nakamura:2010zzi} for the CKM ($V_{\rm CKM}$) and PMNS 
($V_{\rm PMNS}$) matrices as $V_{\rm CKM}\equiv V_{uL}^\dagger V_{dL}$ and $V_{\rm 
PMNS}\equiv V_{eL}^\dagger V_\nu D_p$, respectively, where $V_{\ast L}$ are unitary 
matrices diagonalizing Yukawa coupling as $V_{\ast L}^\dagger y_\ast V_{\ast 
R}=y_\ast^{\rm diag}$, and $D_p$ is a diagonal phase matrix as 
$D_p\equiv\mbox{Diag}\{e^{i\rho},e^{i\sigma},1\}$. The light (active) neutrinos are 
diagonalized as $V^TM_\nu V=M_\nu^{\rm diag}\equiv\mbox{Diag}\{m_1,m_2,m_3\}$ with 
$M_\nu\equiv\kappa v_u^2$. Notice that two Majorana phases in $D_p$ are included 
in the PMNS matrix. 

RGE of $\kappa$ is given by~\cite{Chankowski:1993tx}
 \begin{eqnarray}
  16\pi^2\frac{d\kappa}{dt}
   &=&6\left[-\frac{1}{5}g_1^2-g_2^2+\mbox{Tr}\left(y_u^\dagger y_u\right)
      \right]\kappa \nonumber \\
   & & +\left[\left(y_ey_e^\dagger\right)\kappa
              +\kappa\left(y_ey_e^\dagger\right)^T\right], \label{RGE}
 \end{eqnarray}
where $g_i$s are gauge coupling constants. We can show the PMNS matrix at a 
high energy scale, $\Lambda$, through the neutrino mass matrix at $\Lambda$, 
$M_\nu(\Lambda)=\kappa(\Lambda)v_u^2$, by use of eq.(\ref{RGE}). The 
$M_\nu(\Lambda)$ is given by 
$M_\nu(\Lambda)=IM_\nu(\Lambda_{\rm EW})I$~\cite{Ellis:1999my,Haba:1998fb,Haba:1999fk,Haba:1999xz}, where $\Lambda_{\rm EW}$ is a low energy (electroweak) 
scale and $I\equiv\mbox{Diag}\{\sqrt{I_e},\sqrt{I_\mu},\sqrt{I_\tau}\}$. Here, 
$I_\alpha$s ($\alpha=e,\mu,\tau$) denote quantum corrections, which are defined
 by $I_\alpha\equiv\mbox{exp}\left[\frac{1}{8\pi}\int_{t_\Lambda(\equiv\ln\Lambda)}^{t_{\rm EM}(\equiv\ln\Lambda_{\rm EW})}y_\alpha^2dt\right]$. A dominant 
effect of the quantum corrections comes from $y_\tau$, and we define small 
parameters as $\epsilon_{e(\mu)}\equiv1-\sqrt{\frac{I_\tau}{I_{e(\mu)}}}$. 
Typical values of $\epsilon_{e(\mu)}$ have been given in~\cite{Haba:1999fk}, 
and we take a region $10^{-3}\lesssim\epsilon_{e(\mu)}\lesssim0.1$, which 
corresponds to $\mathcal{O}(10)\lesssim\tan\beta\lesssim\mathcal{O}(50)$ with 
$\kappa^{-1}\geq10^{13}$ GeV. In the following analyses, we take a good 
approximation of $\epsilon\equiv\epsilon_e=\epsilon_\mu$. Then, the 
$M_\nu(\Lambda)$ is given by
 \begin{eqnarray}
  M_\nu(\Lambda)=
   \left(
    \begin{array}{ccc}
     m_{11}             & m_{12}             & m_{13}(1+\epsilon)  \\
     m_{21}             & m_{22}             & m_{23}(1+\epsilon)  \\
     m_{31}(1+\epsilon) & m_{32}(1+\epsilon) & m_{33}(1+2\epsilon)
    \end{array}
   \right),
 \end{eqnarray}
and
 \begin{align}
  &(M_\nu(\Lambda_{\rm EW}))_{ij} \nonumber \\
  &\equiv m_{ij}
  =(V_{\rm PMNS}^\ast(\Lambda_{\rm EW})
   M_\nu^{\rm diag}(\Lambda_{\rm EW})
   V_{\rm PMNS}^T(\Lambda_{\rm EW}))_{ij}. \label{neutrinoMM}
 \end{align}
Note that we take a diagonal basis of charged lepton Yukawa matrix.
 
\begin{figure}
\begin{tabular}{cc}
(a) & (d) \\
\includegraphics[scale=0.16]{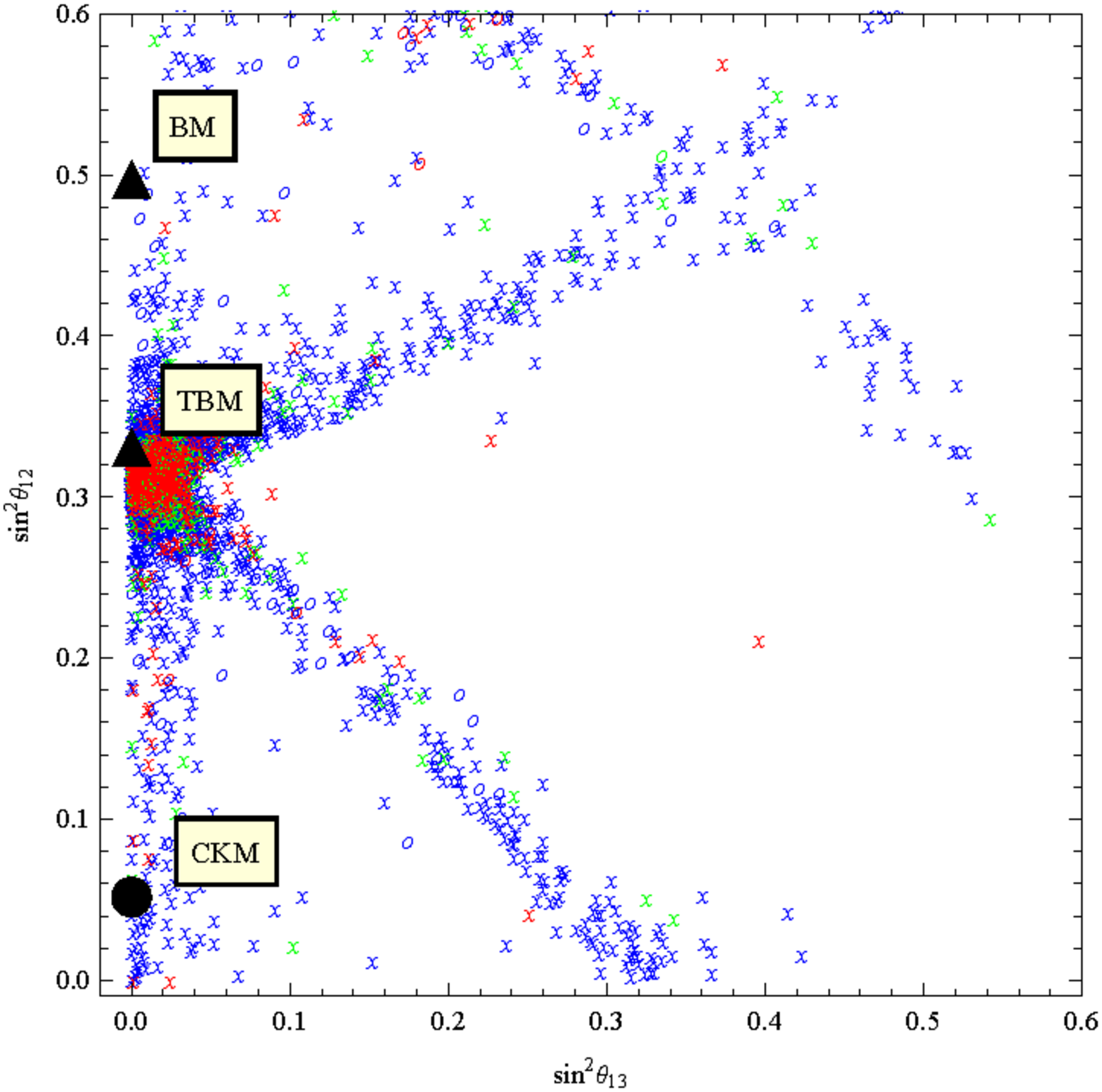} & 
\includegraphics[scale=0.16]{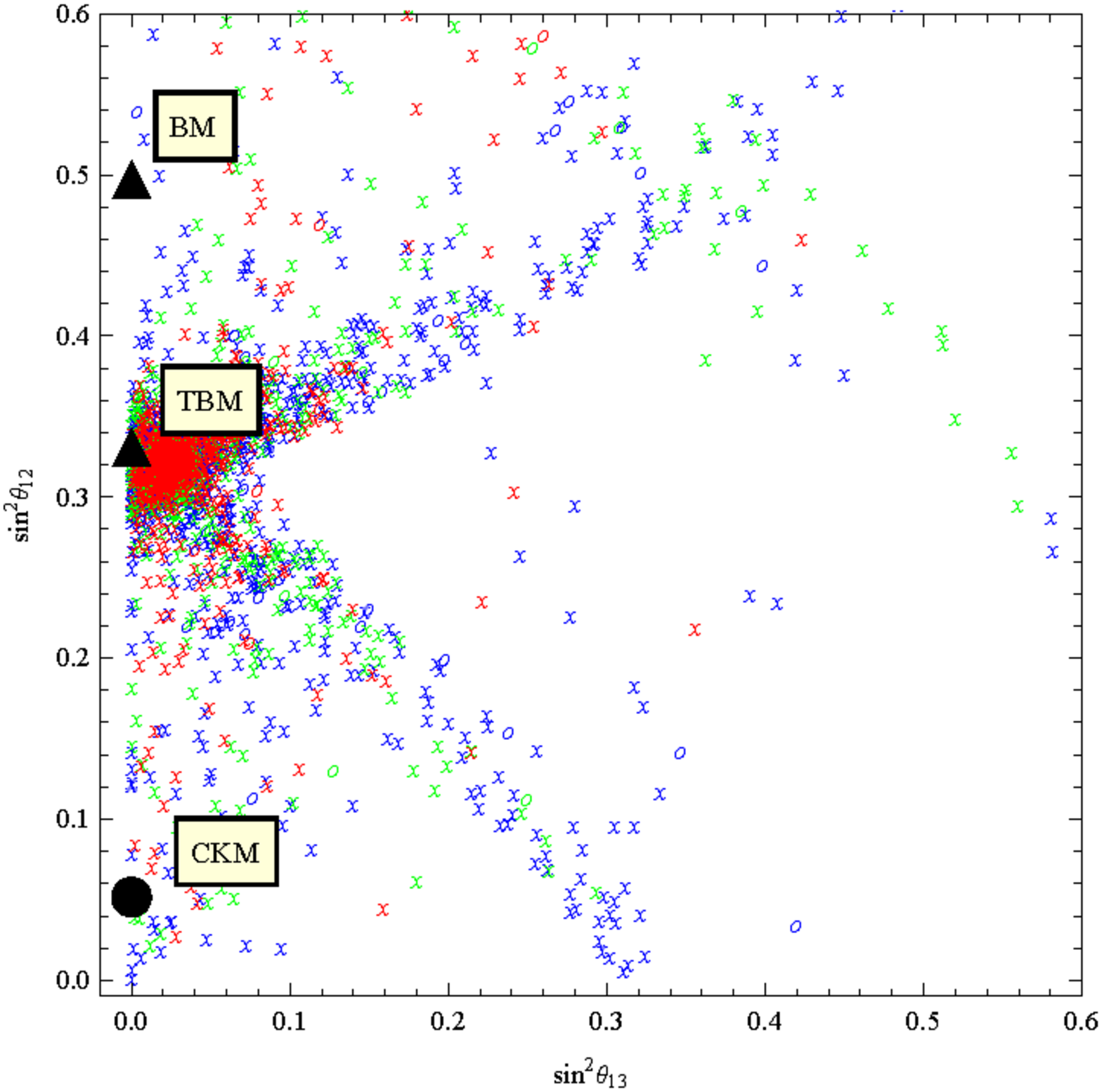} \\
(b) & (e) \\
\includegraphics[scale=0.16]{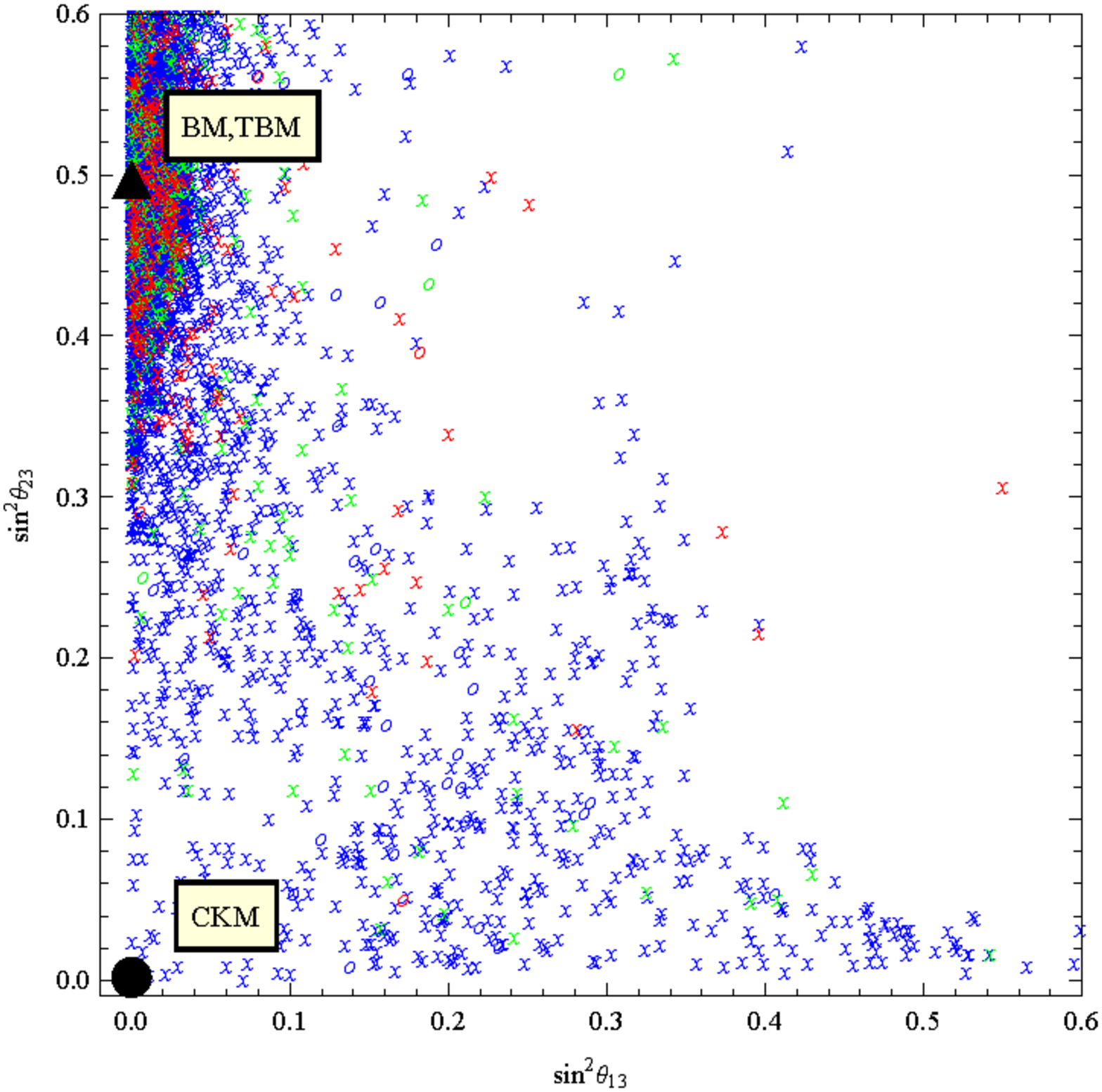} & 
\includegraphics[scale=0.16]{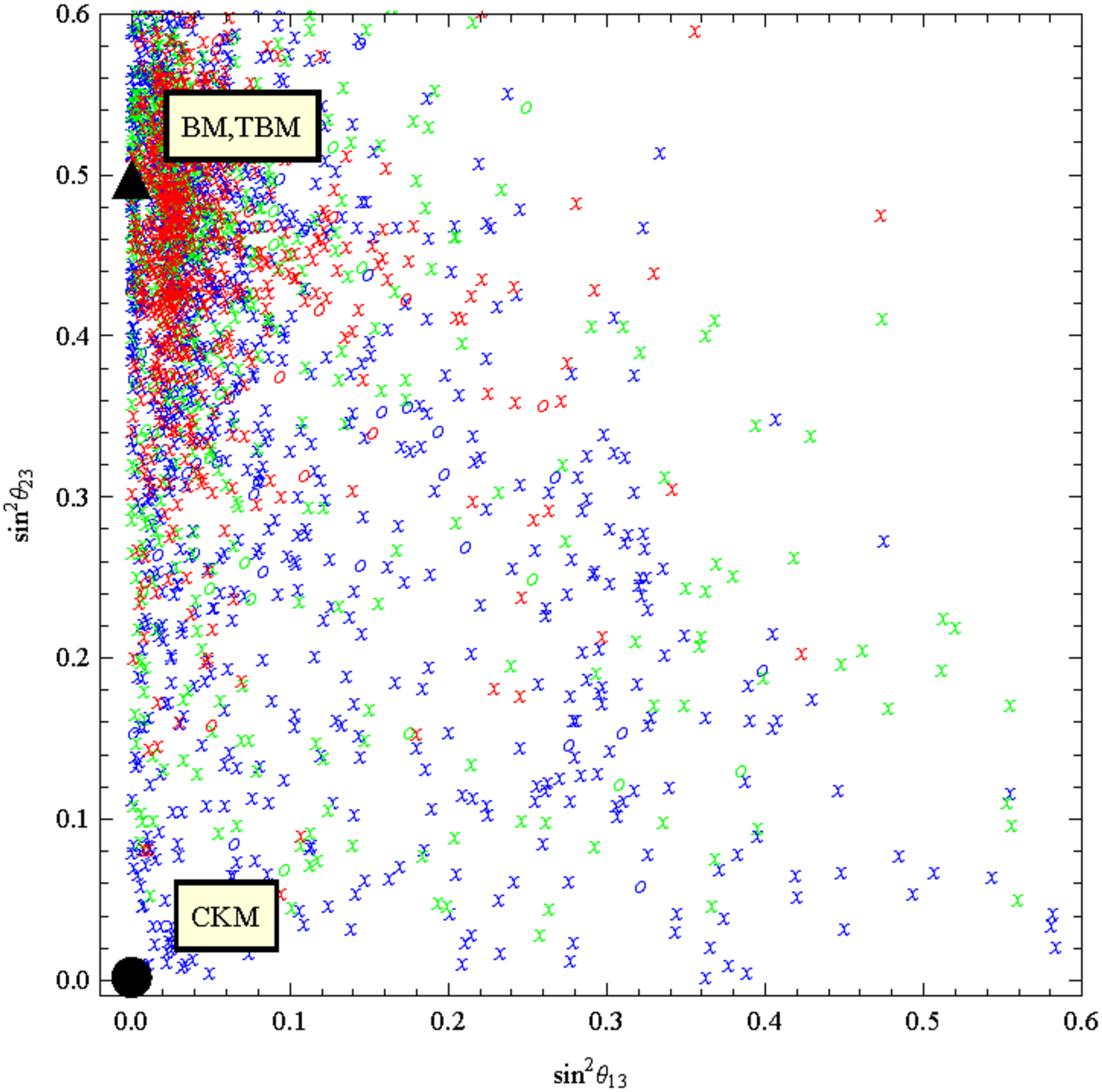} \\
(c) & (f) \\
\includegraphics[scale=0.16]{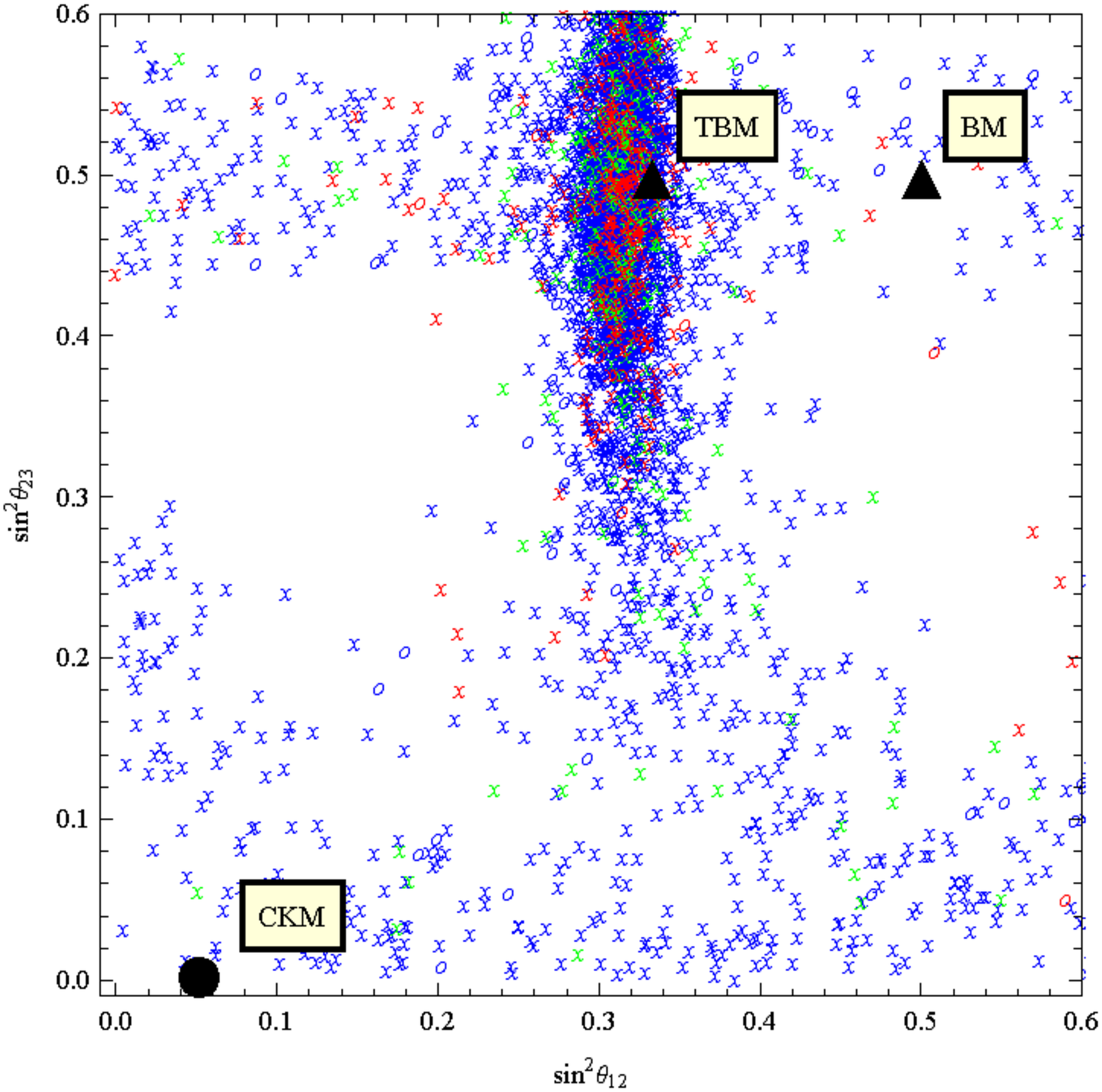} &
\includegraphics[scale=0.16]{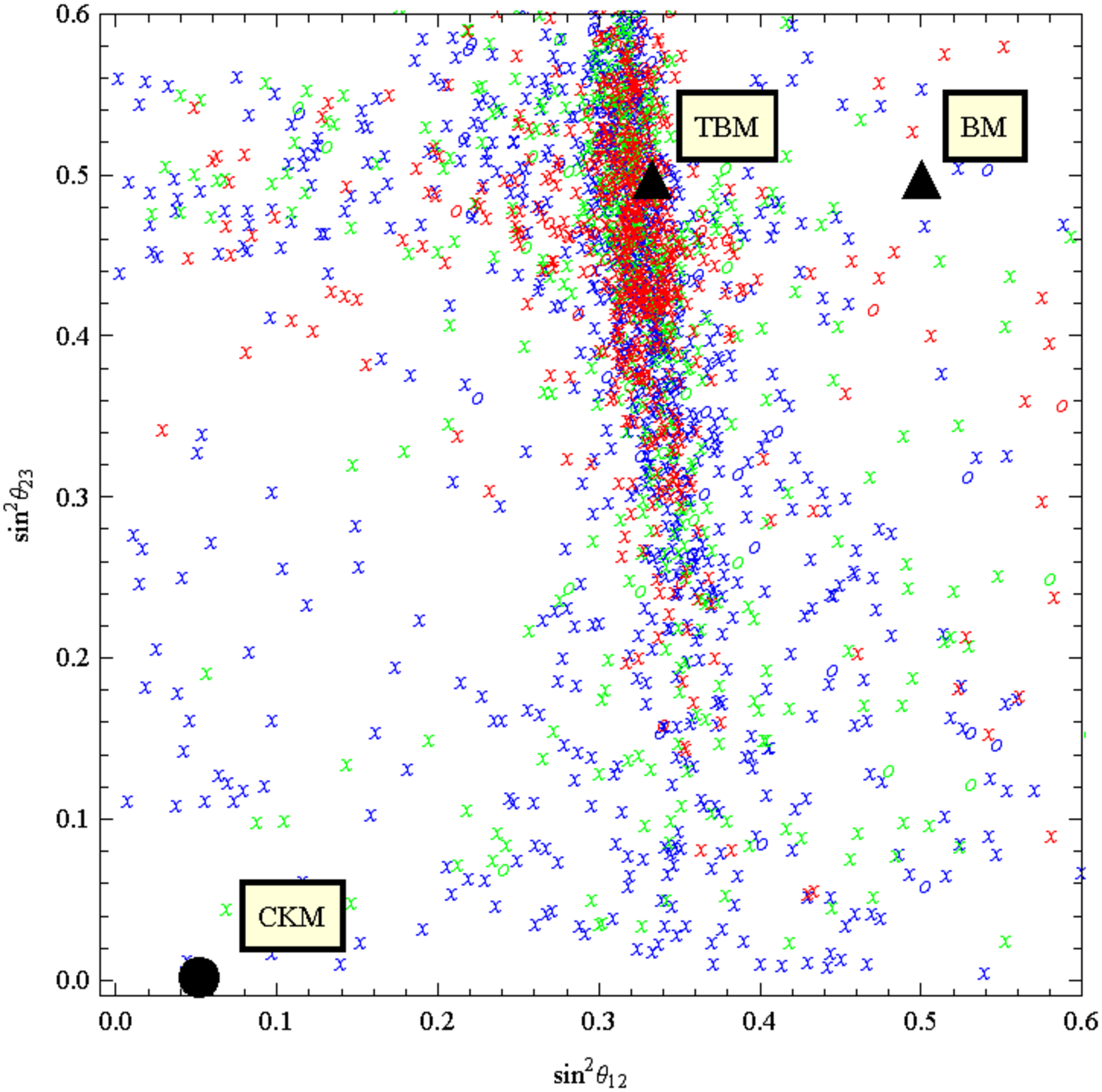} \\
\end{tabular}
\caption{The PMNS mixing angle at $\Lambda=10^{14}$ GeV for the NH case. Red, 
green and blue plots show large hierarchy ($\sqrt{|\Delta m_{32}^2|+\Delta 
m_{21}}\leq m_3<0.1$ eV), weak degenerate ($0.1\mbox{ eV}\leq m_3<0.15$ eV) and 
strong degenerate ($0.15\mbox{ eV}\leq m_3\leq0.2$ eV) cases, respectively.} 
\label{fig1}
\end{figure}

\begin{figure}
\begin{tabular}{cc}
(a) & (d) \\
\includegraphics[scale=0.16]{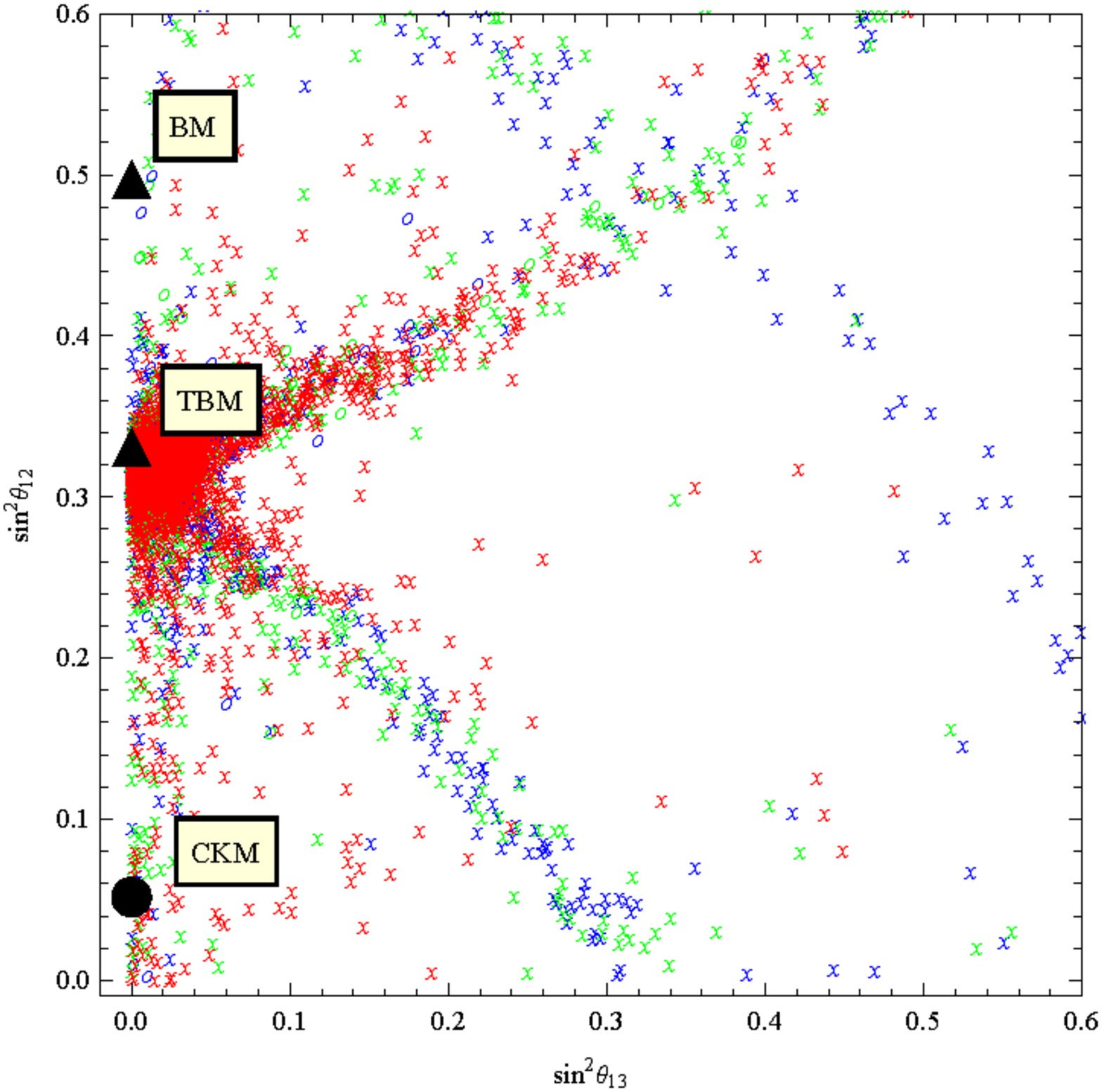} & 
\includegraphics[scale=0.16]{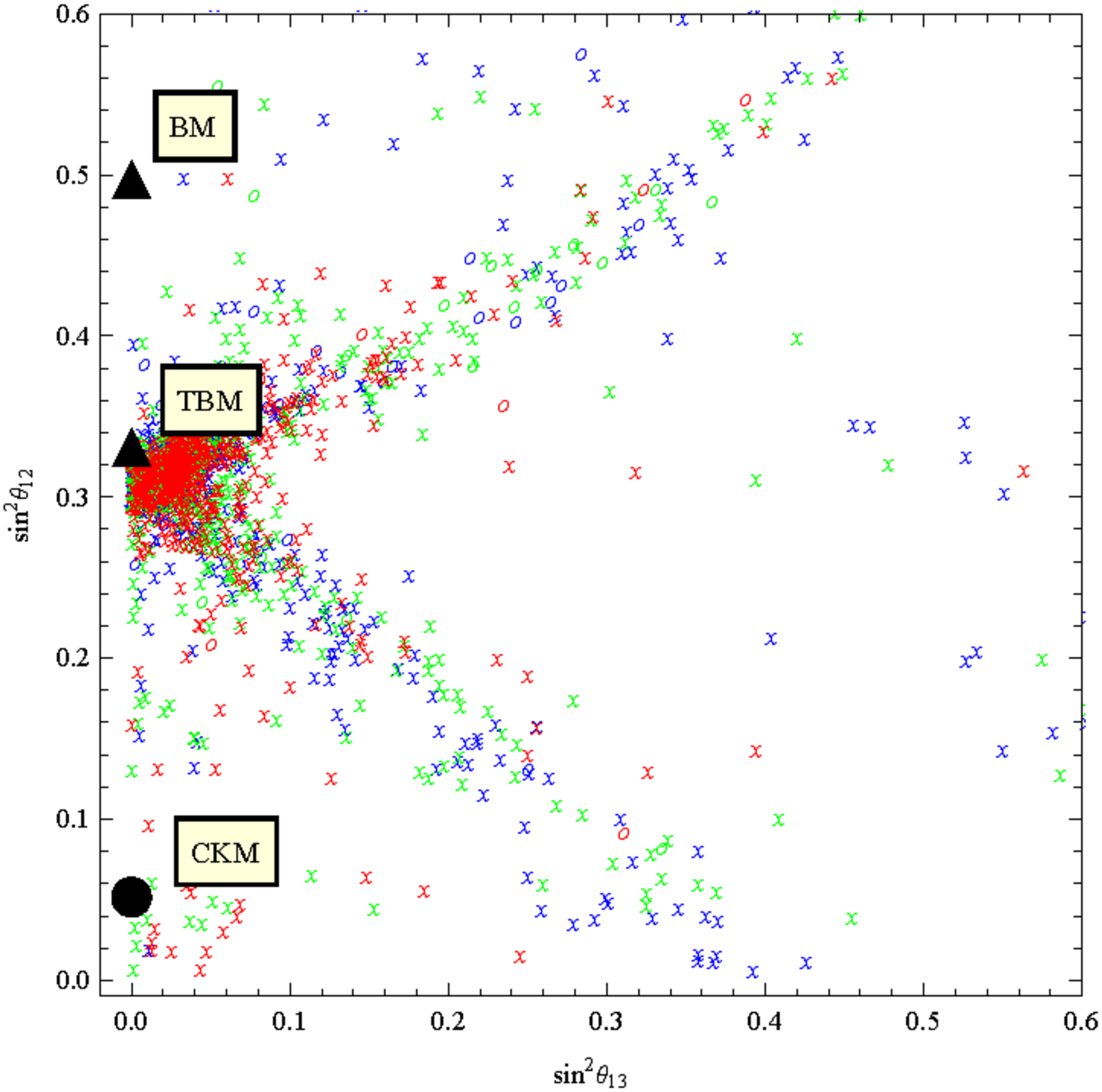} \\
(b) & (e) \\
\includegraphics[scale=0.16]{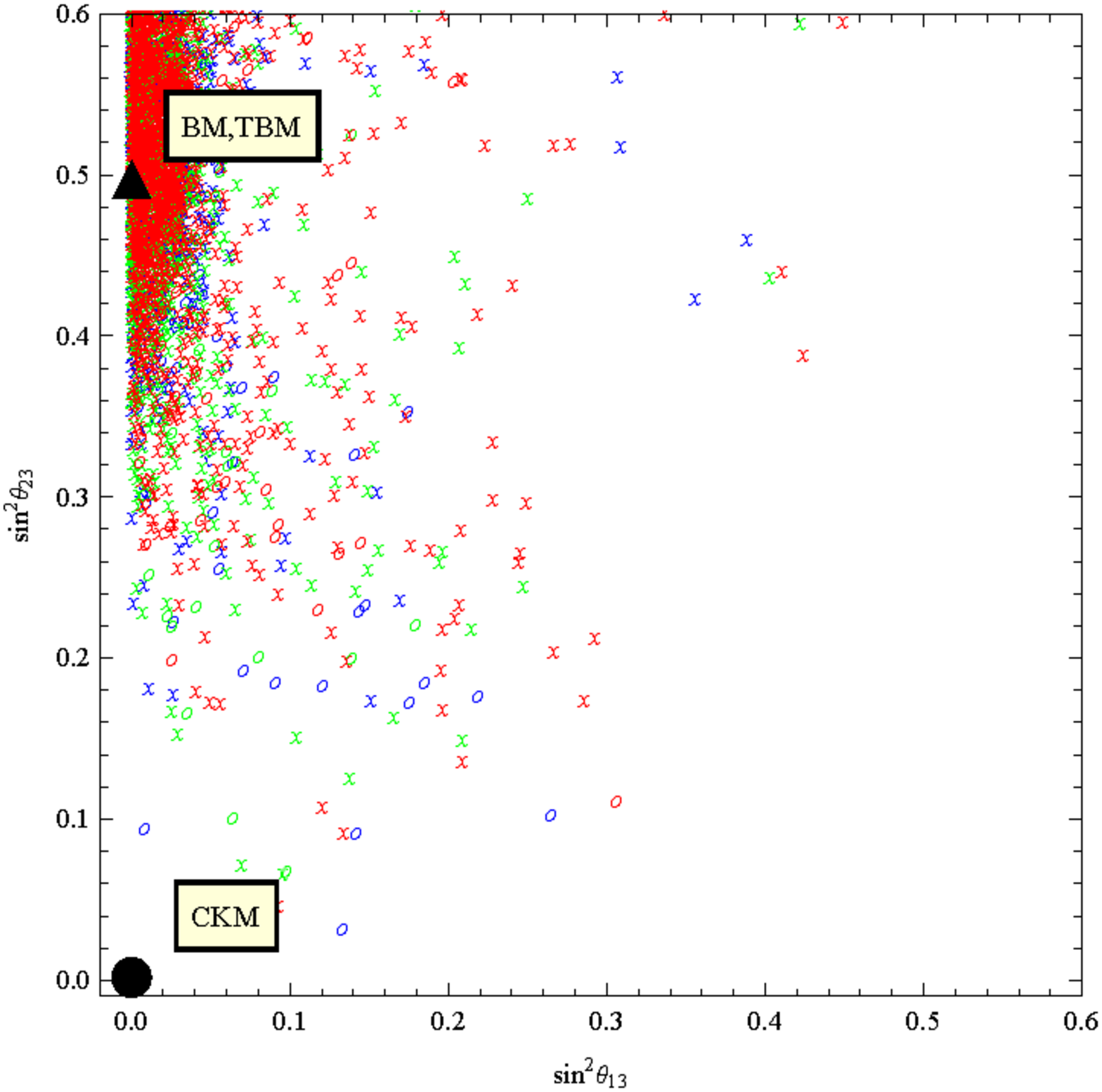} & 
\includegraphics[scale=0.16]{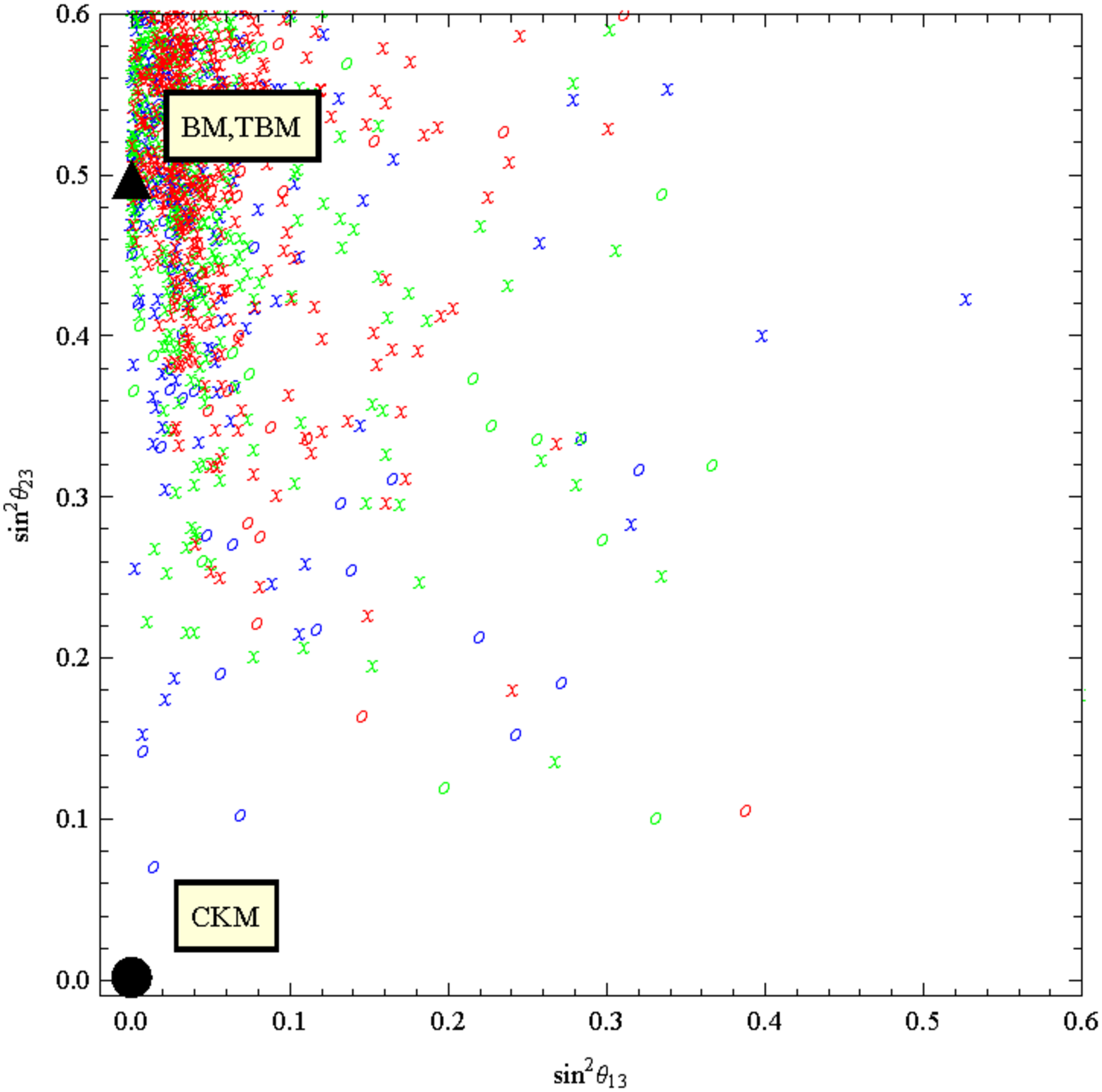} \\
(c) & (f) \\
\includegraphics[scale=0.16]{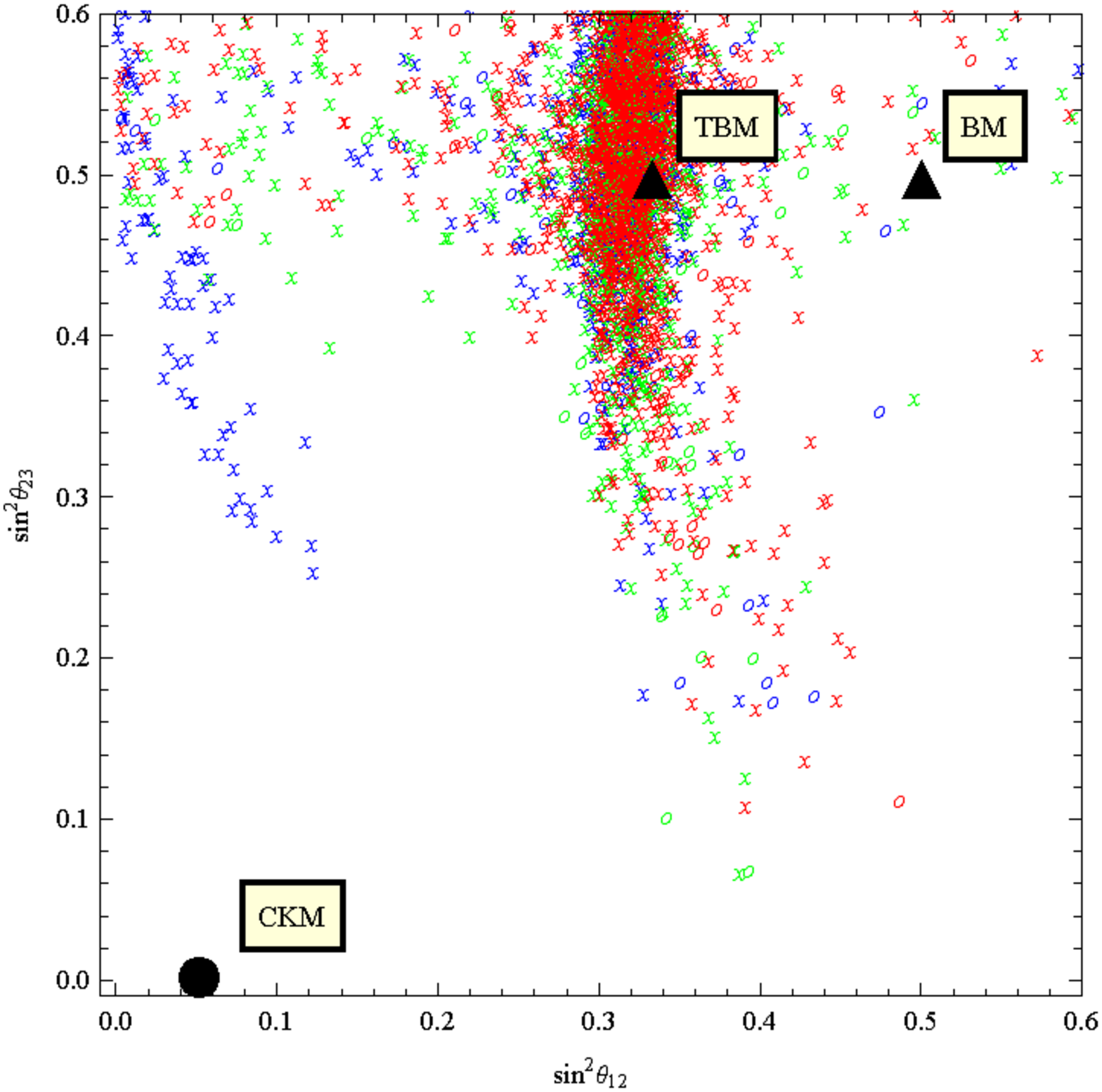} &
\includegraphics[scale=0.16]{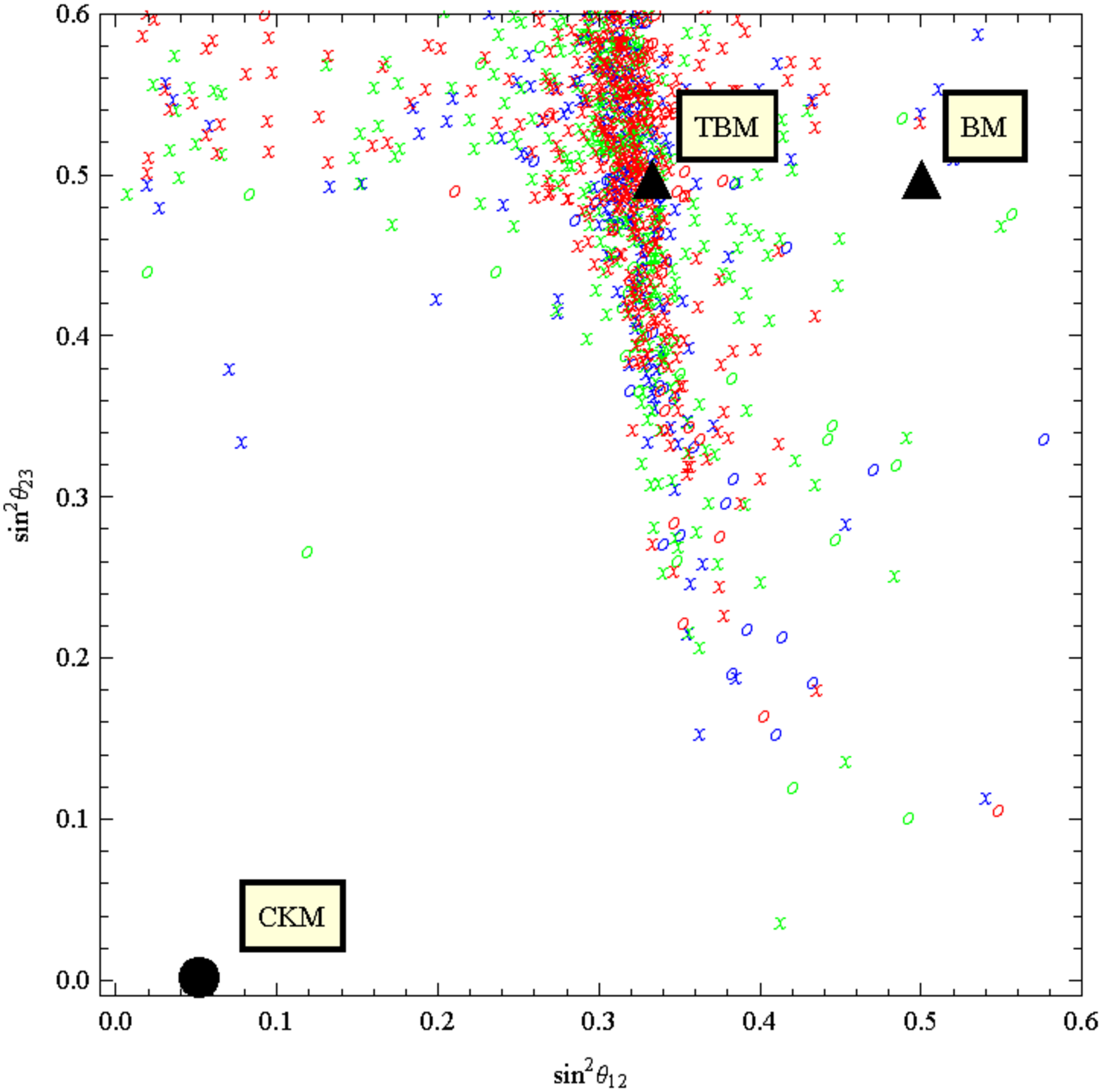} \\
\end{tabular}
\caption{The same figures as Fig.1 for the IH case. Red, green and blue plots 
show large hierarchy ($\sqrt{|\Delta m_{32}^2|+\Delta m_{21}}\leq m_2<0.1$ eV),
 weak degenerate ($0.1\mbox{ eV}\leq m_2<0.15$ eV) and strong degenerate 
($0.15\mbox{ eV}\leq m_2\leq0.2$ eV) cases, respectively.} 
\label{fig2}
\end{figure}

Now let us investigate effects of radiative corrections for 
 the PMNS mixing angles. 
Numerical results are shown in Figs.~\ref{fig1} and \ref{fig2}. We have 
performed scatter plots with the following input parameters. 

For the mass spectra of light neutrinos, we take two types of neutrino mass 
ordering, normal hierarchy (NH) $m_1<m_2<m_3$ and inverted hierarchy (IH) 
$m_3<m_1<m_2$, since the neutrino oscillation experiments determine only two 
mass squared differences, $\Delta m_{21}^2\equiv |m_2|^2-|m_1|^2$ and 
$|\Delta m_{32}^2|\equiv\left||m_3|^2-|m_2|^2\right|$. At the 
$\Lambda_{\rm EW}$ scale, the NH case suggests 
 \begin{eqnarray}
  m_1(\Lambda_{\rm EW})
   &=& \sqrt{m_3^2(\Lambda_{\rm EW})-|\Delta m_{32}^2|-\Delta m_{21}^2}, \\
  m_2(\Lambda_{\rm EW})
   &=& \sqrt{m_3^2(\Lambda_{\rm EW})-|\Delta m_{32}^2|},
 \end{eqnarray}
while the IH case does 
 \begin{eqnarray}
  m_1(\Lambda_{\rm EW})
   &=& \sqrt{m_2^2(\Lambda_{\rm EW})-\Delta m_{21}^2}, \\
  m_3(\Lambda_{\rm EW})
   &=& \sqrt{m_2^2(\Lambda_{\rm EW})-|\Delta m_{32}^2|-\Delta
   m_{21}^2}. 
 \end{eqnarray}
We have taken $\sqrt{|\Delta m_{32}^2|+\Delta 
m_{21}^2}\leq m_{3(2)}(\Lambda_{\rm EW})\leq0.2$ eV for NH (IH) with 
 \begin{eqnarray}
  \Delta m_{21}^2 &=& 7.62\times10^{-5}\mbox{ eV}^2, \\
  |\Delta m_{32}^2| &=& 2.53(2.40)\times10^{-3}\mbox{ eV}^2,
 \end{eqnarray}
which are the best fit values of experimentally observed neutrino mass squared 
differences~\cite{tortola}. The magnitude of 0.2 eV is consistent with 
cosmological bounds on sum of neutrino masses (see e.g.~\cite{Komatsu:2010fb}). 
The mixing angles at $\Lambda_{\rm EW}$ are taken by 
 \begin{align}
  &0.303\leq\sin^2\theta_{12}\leq0.335, \\
  &0.44(0.46)\leq\sin^2\theta_{23}\leq0.57(0.58), \\
  &0.022(0.023)\leq\sin^2\theta_{13}\leq0.029(0.030), \label{BCPMNS}
 \end{align}
{}from experimental results at $1\sigma$ level for the NH (IH) 
case~\cite{tortola}. Notice that $m_3(\Lambda_{\rm EW})$ ($m_2(\Lambda_{\rm EW})$) 
is a free parameter in our analyses for the NH (IH) case, and it is related to 
the magnitude of degeneracy, i.e., a larger $m_3(\Lambda_{\rm EW})$ 
($m_2(\Lambda_{\rm EW})$) stands for a stronger degeneracy. When the degeneracy 
becomes stronger, the mixing angles can change drastically.  

The effects of quantum correction described by $\epsilon$ have been taken as 
$10^{-3}\leq\epsilon\leq0.1$. In the figures, the ``$o$'' and ``$\chi$'' 
markers show relatively small $\epsilon$ ($10^{-3}\leq\epsilon<0.01$) and large
 one ($0.01\leq\epsilon\leq0.1$), respectively. Note that $\epsilon$ is also 
a free parameter in our analyses, which is determined once values of 
$\tan\beta$ and $\Lambda$ are fixed. The scatter plots in Figs.~\ref{fig1} and 
\ref{fig2} denote the PMNS mixing angles for NH and IH cases in a typical high 
energy scale of $\Lambda=10^{14}$ GeV, respectively. We analyze separately 
whether all CP-phases are relatively large 
$\pi/4\leq|\delta^l|,|\rho|,|\sigma|$ ((a)-(c)) or small 
$0\leq|\delta^l|,|\rho|,|\sigma|<\pi/4$ ((d)-(f)). The CKM mixing angles at 
$\Lambda=10^{14}$ GeV~\cite{Ross:2007az} is shown in each figure by big black 
dot. 

Now let us go back our starting point, ``Can the GUFM be really possible?'' For
 the NH case, there definitely exist regions of the GUFM, where all the PMNS 
mixing angles are equal to the corresponding CKM mixing angles, 
$\theta_{ij}^l=\theta_{ij}^q$ as shown by blue markers in Figs.~\ref{fig1}. 
Notice also that the green and red markers cannot reach the CKM point (e.g. see
 Figs.~\ref{fig1} (c) and (f)). Large hierarchy and small degenerate cases 
cannot realize GUFM because of the stabilities of mixing angle $\theta_{23}$. 
Therefore, the GUFM can be achieved in a case of strong degenerate neutrino 
mass spectrum through quantum corrections. As for CP-phases, the GUFM is easily
 realized when CP-phases are large. We can see it by comparing Figs.~\ref{fig1}
 (a)-(c) with Figs.~\ref{fig1} (d)-(f). As for the largeness of quantum 
corrections, the strong degenerate case (blue plot) can realize CKM angles even
 when the quantum effects are relatively small since blue ``$o$'' markers in 
Fig.~\ref{fig1} (b) or (c) really exist on CKM point. Numerically, we can see 
that $0.005\lesssim\epsilon$, which corresponds to $10\lesssim\tan\beta$, is 
enough for the realization of the GUFM in the NH degenerate case. 

On the other hand, Figs.~\ref{fig2} show that the IH case cannot realize the 
GUFM. It is because $\theta_{23}$ becomes too large at $\Lambda=10^{14}$ GeV. 
Thus, we can conclude that the strong degenerate NH mass spectrum can achieve 
the GUFM in a region of $0.002(0.005)\lesssim\epsilon$ with the large (small) 
CP-phases case, which corresponds to 
$\tan\beta\simeq10(15)$~\cite{Haba:1999fk}. This situation is summarized by 
``CKM'' in Tab.~\ref{tab1}. Additionally, ``CKM'' in Tab.~\ref{tab2} shows 
cases of different combinations of CP-phases, such as one of three is small 
(large) and others are large (small). In these cases, numerical results are not
 so changed,  and we can conclude that the most important key for the 
realization of GUFM is not CP-phases but strong degeneracy. 

We give some comments on our the results. First, our results are consistent with
 ones of~\cite{Mohapatra:2003tw}. The work of~\cite{Mohapatra:2003tw} utilized 
relatively large $m_3(\geq0.17)$ and $\tan\beta$(=55), which are favor for the 
realization of GUFM as we have shown. These values of $m_3$ and $\tan\beta$ are 
inputs in~\cite{Mohapatra:2003tw} while we have scanned over $m_3$ and 
$\tan\beta$, and we have successfully obtained lower bounds on $m_3$ and 
$\tan\beta$ for the GUFM in this work. We have also shown that the GUFM cannot 
be realized in the IH case. Second, there generically exist threshold effects 
for neutrino masses~\cite{Chun}. We did not take care of such effects because it
 was shown in~\cite{Mohapatra:2005gs} that the threshold corrections have 
negligible effects on the mixing angles, and thus size of the effects is 
sufficient to have concordance between the GUFM model and experimental results 
of neutrino oscillation.

We also comment on a correlative mixing pattern, 
$\theta_{12}+\theta_{23}+\theta_{13}=\pi/2$, at a low energy scale. Even when 
we change value of $\theta_{13}$ as keeping the relation 
$\theta_{12}+\theta_{23}+\theta_{13}=\pi/2$ within the experimentally allowed 
values, the main results given in Tab.~\ref{tab1} are not changed.

\begin{table}
\begin{center}
\begin{tabular}{c||c|c}
\hline
 & \multicolumn{2}{|c}{NH ($m_1<m_2<m_3$)}  \\
\hline
 & $\frac{\pi}{4}\leq|\delta^l|,|\rho|,|\sigma|$ & $0\leq|\delta^l|,|\rho|,|\sigma|<\frac{\pi}{4}$ \\
\hline\hline
 & $\bigcirc$ & $\bigcirc$ \\ 
CKM & $0.1\mbox{ eV}\lesssim m_3$ & $0.15\mbox{ eV}\lesssim m_2$ \\
 & $10\lesssim\tan\beta$ & $15\lesssim\tan\beta$ \\
\hline
BM & \begin{tabular}{@{}c@{}} $\bigcirc$ \\ $8\lesssim\tan\beta$ \end{tabular} & $\times$ \\
\hline
TBM & $\bigcirc$ &$\bigcirc$ \\
\hline\hline
Figs. & Figs.~\ref{fig1} (a)-(c) & Figs.~\ref{fig1} (d)-(f) \\
\hline
\end{tabular}
\begin{tabular}{c||c|c}
\hline	
& \multicolumn{2}{|c}{IH ($m_3<m_1<m_2$)} \\
\hline
& $\frac{\pi}{4}\leq|\delta^l|,|\rho|,|\sigma|$ & $0\leq|\delta^l|,|\rho|,|\sigma|<\frac{\pi}{4}$ \\
\hline
CKM & $\times$ & $\times$ \\
\hline
BM & \begin{tabular}{@{}c@{}} $\bigcirc$ \\ $10\lesssim\tan\beta$ \end{tabular} & $\times$ \\
\hline
TBM & $\bigcirc$ & $\bigcirc$ \\
\hline\hline
Figs. & Figs.~\ref{fig2} (a)-(c) & Figs.~\ref{fig2} (d)-(f)\\
\hline
\end{tabular}
\end{center}
\caption{This is the summary of main results. 
$\bigcirc$ ($\times$) means that the corresponding mixing angles can (not) be 
realized at high energy scale.}
\label{tab1}
\end{table}
\begin{table}
\begin{center}
\begin{tabular}{ll||c|c|c}
\hline
 & & CKM & BM & TBM \\
\hline\hline
$\frac{\pi}{4}\leq|\delta^l|,|\rho|$, & $0\leq|\sigma|<\frac{\pi}{4}$ & $\bigcirc$ ($\times$) & $\times$ & $\bigcirc$ \\
\hline
$\frac{\pi}{4}\leq|\delta^l|,|\sigma|$, & $0\leq|\rho|<\frac{\pi}{4}$ & $\bigcirc$ ($\times$) & $\times$ & $\bigcirc$ \\
\hline
$\frac{\pi}{4}\leq|\rho|,|\sigma|$, & $0\leq|\delta^l|<\frac{\pi}{4}$ & $\bigcirc$ ($\times$) & $\bigcirc$ & $\bigcirc$ \\
\hline
$\frac{\pi}{4}\leq|\delta^l|$, & $0\leq|\rho|,|\sigma|<\frac{\pi}{4}$ & $\bigcirc$ ($\times$) & $\times$ & $\bigcirc$ \\
\hline
$\frac{\pi}{4}\leq|\rho|$, & $0\leq|\delta^l|,|\sigma|<\frac{\pi}{4}$ & $\bigcirc$ ($\times$) & $\times$ & $\bigcirc$ \\
\hline
$\frac{\pi}{4}\leq|\sigma|$, & $0\leq|\delta^l|,|\rho|<\frac{\pi}{4}$ & $\bigcirc$ ($\times$) & $\times$ & $\bigcirc$ \\
\hline
\end{tabular}
\end{center}
\caption{The realizations of CKM, BM, and TBM in cases of various combinations 
of CP-phases for the NH (IH) case.}
\label{tab2}
\end{table}

\vspace{3mm}
Finally, although it is nothing to do with the GUFM, we comment on bi-maximal 
($\sin^2\theta_{12}=\sin^2\theta_{23}=1/2$ and $\sin^2\theta_{13}=0$) and 
tri-bimaximal ($\sin^2\theta_{12}=1/3$, $\sin^2\theta_{23}=1/2$, and 
$\sin^2\theta_{13}=0$) mixing angles just for reference, which are shown by big
 black triangles in Figs.~\ref{fig1} and \ref{fig2}. We can find regions where 
all the PMNS mixing angles at $\Lambda=10^{14}$ GeV are close to the bi-maximal
 and tri-bimaximal~\cite{TBM} points in Figs.~\ref{fig1} (a)-(c) and 
Figs.~\ref{fig2} (a)-(c). For the BM mixing, $0.0015(0.002)\lesssim\epsilon$ is 
required for NH (IH) case, which corresponds to 
$\tan\beta\simeq8(10)$~\cite{Haba:1999fk}. And, the BM mixing cannot be realized
 in small CP-phases cases both for NH and IH cases. In different combinations of
 CP-phases, the BM cannot be realized unless $\pi/4\leq|\rho|,|\sigma|$ for both
 NH and IH cases. These mean the largeness of $|\rho|$ and $|\sigma|$ is 
important for the realization of BM at the high energy scale (see 
Tab.~\ref{tab2}). On the other hand, the TBM mixing angles at the high energy 
are allowed for all cases (NH/IH and large/small CP-phases (see 
Tab.~\ref{tab2})). All figures show that the TBM is easy to be realized at high 
energy scale with relatively small quantum effects (``$o$'' marker), since the 
TBM fits the PMNS mixing angles well at the low energy scale. 


\vspace{3mm}
We have investigated whether the GUFM is possible or not in the framework of 
the MSSM. We have found the GUFM is really possible when neutrino has 
degenerated NH spectrum with $0.1$ eV $\lesssim m_3$ through the quantum 
corrections, $0.005\lesssim\epsilon$ ($10\lesssim\tan\beta$). We have also 
investigated the possibility that BM and TBM mixing angles are realized at the 
high energy scale. 
 
\subsection*{Acknowledgment}

This work is partially supported by Scientific Grant by Ministry of Education 
and Science, Nos. 00293803, 20244028, 21244036, and 23340070. The work of R.T. 
is supported by Research Fellowships of the Japan Society for the Promotion of 
Science for Young Scientists.


\end{document}